\documentclass[manuscript,,authorversion,nonacm]{acmart}

\AtBeginDocument{%
  }

\usepackage{soul}
\usepackage{xcolor}

\usepackage{subcaption}

\usepackage{multirow}
\usepackage{caption}

\setcopyright{acmcopyright}
\copyrightyear{2018}
\acmYear{2018}
\acmDOI{XXXXXXX.XXXXXXX}

\acmConference[AHs '23]{The Augmented Humans (AHs) International Conference}{March 12-14, 2023}{Glasgow, United Kingdom}

\acmBooktitle{Augmented Humans International Conference (AHs '23), March 12-14, 2023, Glasgow, United Kingdom}

\acmPrice{15.00}
\acmISBN{978-1-4503-XXXX-X/18/06}

\begin{document}

\title[Exoskeleton for the Mind]{Exoskeleton for the Mind: Exploring Strategies Against Misinformation with a Metacognitive Agent}

\author{Yeongdae Kim}
\orcid{0000-0001-5346-0041}
\affiliation{%
  \institution{Tokyo Institute of Technology}
  \streetaddress{2 Chome-12-1 Ookayama}
  \city{Meguro}
  \state{Tokyo}
  \country{Japan}
  \postcode{152-8550}
}
\email{kim.y.ah@m.titech.ac.jp}

\author{Takane Ueno}
\orcid{0000-0001-6626-7945}
\affiliation{%
  \institution{Tokyo Institute of Technology}
  \streetaddress{2 Chome-12-1 Ookayama}
  \city{Meguro}
  \state{Tokyo}
  \country{Japan}
  \postcode{152-8550}
}\email{ueno.t.ao@m.titech.ac.jp}

\author{Katie Seaborn}
\orcid{0000-0002-7812-9096}
\affiliation{%
  \institution{Tokyo Institute of Technology}
  \streetaddress{2 Chome-12-1 Ookayama}
  \city{Meguro}
  \state{Tokyo}
  \country{Japan}
  \postcode{152-8550}
}\email{seaborn.k.aa@m.titech.ac.jp}

\author{Hiroki Oura}
\orcid{0000-0001-5111-7864}
\affiliation{%
  \institution{Tokyo University of Science}
  \streetaddress{1 Chome-3 Kagurazaka}
  \city{Shinjuku}
  \state{Tokyo}
  \country{Japan}
  \postcode{162-8601}
}\email{houra@rs.tus.ac.jp}

\author{Jacqueline Urakami}
\orcid{0000-0002-2866-0807}
\affiliation{%
  \institution{Tokyo Institute of Technology}
  \streetaddress{2 Chome-12-1 Ookayama}
  \city{Meguro}
  \state{Tokyo}
  \country{Japan}
  \postcode{152-8550}
}\email{urakami.j.aa@m.titech.ac.jp}

\author{Yuto Sawa}
\affiliation{%
  \institution{Tokyo Institute of Technology}
  \streetaddress{2 Chome-12-1 Ookayama}
  \city{Meguro}
  \state{Tokyo}
  \country{Japan}
  \postcode{152-8550}
}\email{sawa.y.ac@m.titech.ac.jp}
\renewcommand{\shortauthors}{Kim et al.}

\begin{abstract}
  Misinformation is a global problem in modern social media platforms with few solutions known to be effective. Social media platforms have offered tools to raise awareness of information, but these are closed systems that have not been empirically evaluated. Others have developed novel tools and strategies, but most have been studied out of context using static stimuli, researcher prompts, or low fidelity prototypes. We offer a new anti-misinformation agent grounded in theories of metacognition that was evaluated within Twitter. We report on a pilot study (n=17) and multi-part experimental study (n=57, n=49) where participants experienced three versions of the agent, each deploying a different strategy. We found that no single strategy was superior over the control. We also confirmed the necessity of transparency and clarity about the agent's underlying logic, as well as concerns about repeated exposure to misinformation and lack of user engagement.
\end{abstract}



\maketitle

\section*{Citation}

\noindent
Yeongdae Kim, Takane Ueno, Katie Seaborn, Hiroki Oura, Jacqueline Urakami, and Yuto Sawa. 2023. Exoskeleton for the Mind: Exploring Strategies Against Misinformation with a Metacognitive Agent. In \emph{Proceedings of the Augmented Humans International Conference 2023 (AHs '23)}. Association for Computing Machinery, New York, NY, USA, 209–220. \url{https://doi.org/10.1145/3582700.3582725} \\

\noindent
The final publication is available via ACM at \url{https://dl.acm.org/doi/10.1145/3582700.3582725}.

\section{Introduction}
The digital misinformation age is upon us. Fake news, inaccuracies, both accidental and intended, opinions presented as fact, and misinformed but viral memes are an everyday feature of the social media landscape \cite{allcott2017social,Allcott2019}. 
Social media users are exposed to and have the ability to broadcast content of all kinds with little oversight. 
Professional news sources, published research, and expert opinion are mixed in with, and often indistinguishable form, unverified and disreputable content. 
A combination of proprietary algorithms and human cognitive biases have even created "echo chambers" \cite{colleoni2014echo} rife with misinformation and polarized views. This has led to the present state of affairs: the "infodemic" \cite{zarocostas2020fight}. While the notion of misinformation is not new \cite{hernon1995disinformation}, its scale and spread are a feature of the Internet-connected world in which most people on Earth currently live. 
For example, there is substantial evidence that social media was harnessed by bad actors in order to interfere with the US 2016 election \cite{allcott2017social} and galvanize violent action that led to the 2021 US capitol riot \cite{bleakley2021panic, garrett2021drug}. Misinformation has also been a feature of social media output on the COVID-19 pandemic \cite{Kouzy2020}. 
This has led to urgent calls for ways in which to regulate social media \cite{cusumano2021social}, understand how misinformation influences people \cite{Geeng2020, Allcott2019, Shin2018, Wu2016}, and provide people with tools to fight misinformation in these spaces \cite{Vraga2017, Bautista2021, Walter2020, Mena2020, Ross2018}.

Social media companies 
explored automated and manual content curation, filters, warning labels, and policy updates \cite{niemiec2020covid, stjernfelt2020your}. However, criticism and concerns have been raised about moderation inconsistency and censorship \cite{stjernfelt2020your}, as well as conflicts of interest and a lack of known empirical testing when it comes to company fact-checking tools, such as Meta/Facebook's Third-Party Fact-Checking Program\footnote{\url{https://www.facebook.com/formedia/mjp/programs/third-party-fact-checking}} and Twitter's Community Notes\footnote{\url{https://twitter.github.io/communitynotes}}. 
Another promising direction is 
\emph{third-party} anti-misinformation tools. Previous research has explored a variety of strategies, such as warning labels \cite{bhuiyan2018, hartwig2019, pennycook2018prior}, fact-checking \cite{vijjali2020two, Kim2020, chi2022quantitative}, asking users to pause and reflect \cite{fazio2020, Bago2020, pennycook2021shifting}, and so on.
Yet, 
most of this work has been conducted in lab environments with artificial stimuli, using printouts of doctored social media posts.
Others have relied on quantitative methods, limiting access to richer forms of experiential data, such as observations of how people attend and react to certain content over others. This also limits access to the "why" of behaviour and the meaning of the quantitative results. Also, most work has focused on one strategy in isolation, which makes drawing comparisons among strategies to identify the best strategies very difficult. Finally, several strategies, especially well-studied and widely used strategies like warning labels, have been found ineffective. A different approach to the design and study of anti-misinformation tools in social media may strengthen this crucial area of work and lead to clearer results. 

To this end, we created a new anti-misinformation agent for live social media contexts that deploys a range of strategies informed by theories of metacognition in the digital age \cite{kozyreva2020citizens}. Metacognition means "thinking about one's thinking" \cite{dunlosky2008metacognition}. A "metacognitive agent," then, attempts to raise the user's awareness of their thinking. In naming our agent, we translated a common metaphor from \emph{physical} human augmentation systems to the case of metacognition: \emph{Elemi}, an "\emph{\textbf{\underline{e}}xoske\textbf{\underline{le}}ton for the \textbf{\underline{mi}}nd}."
We aimed to build on previous work by: (i) 
using real content in a live platform known to have a misinformation problem, i.e., Twitter \cite{Allcott2019}; (ii) directly comparing several metacognitive strategies; and (iii) combining quantitative and qualitative forms of inquiry, i.e., mixed methods, to increase clarity on agent and strategy efficacy.
We asked: \emph{Can a metacognitive agent be effective at helping people tackle misinformation within a live social media context?} 
We conducted a pilot study on the basic form of the agent (n=17) and a multi-phased experimental study (n=57 in Session 1 and n=49 in Session 2) comparing three strategies deployed by the agent---\emph{Cover}, \emph{Camouflage}, and \emph{Consider}---against a control. We also evaluated the presence of an \emph{illusory truth effect} \cite{pennycook2018prior} 
by checking for changes in perceptions of content accuracy influenced by exposure in the first session. We used a curated selection of real Tweets
to increase participant comfort and honesty, in response to recent findings on social stigma about misinformation \cite{ren2023beyond}. We aimed to strike a balance between realism and experimental control against the state of our prototype.
We contribute: a proof-of-concept prototype that offers three strategies and was built for a live environment; a methodological approach to evaluating such a tool in situ; and insights into the acceptance, value, behaviour, and user experience (UX) of such a tool for social media users.
We end with implications for designing such tools and directions for future research.

\section{Related Work}
We discuss the tools and strategies being explored to combat misinformation and how our agent builds on this foundation.

\subsection{Anti-Misinformation Tools That Warn Us}
Algorithms can be (re)developed to warn users when they are potentially or actually exposed to false information \cite{bhuiyan2018, sharevski2021misinformation, hartwig2019, Saltz2021, pennycook2018prior}. Bhuiyan et al. \cite{bhuiyan2018} explored the effect of warnings on the perceived credibility of news content in \emph{FeedReflect}, a browser extension for Twitter. Content that contained false information was highlighted or blurred with a warning popup. Findings showed that the warning nudge treatment increased the perceived credibility of the mainstream content while decreasing the credibility of the non-mainstream, misinformed content. However, this work was a small pilot test. Similarly, Sharevski et al. \cite{sharevski2021misinformation} investigated the impact of warning labels in Twitter. They utilized two soft moderation methods: \emph{Warning Cover} and \emph{Warning Tag}. They found that a simple Warning Tag was insufficient to reduce the perceived accuracy of misleading content. Even so, a Warning Cover led users to perceive the accuracy of misleading content negatively, i.e., users trusted content with misinformation less. But they tested it with content that they created, rather than real content. Hartwig and Reuter \cite{hartwig2019} created a browser extension for Twitter called \emph{TrustyTweet}, which used a "white box" approach to inform users about the reasons behind algorithmic fact-checking. The tool attached a warning icon in combination with an explanation when clicked. Findings suggested that the tool was generally considered helpful and maintained the user's sense of autonomy. However, this was a small study and the guidelines need further confirmation. We chose the \emph{Warning Cover} strategy \cite{sharevski2021misinformation} as our first strategy, evaluating it in our pilot study as the \emph{Cover} version of the agent and then iterating its design based on user feedback for the experimental study. We used the "white box" approach \cite{hartwig2019} for the \emph{Camouflage} version of the agent. We built on Bhuiyan et al. \cite{bhuiyan2018} by running a full study with a larger sample of participants.

\subsection{Anti-Misinformation Tools That Ask Us to Pause and Reflect}
In a "pause and reflect" strategy, the user is prompted to stop for a moment and reflect on the content that they are consuming \cite{fazio2020,Bago2020}. Fazio \cite{fazio2020} explored a prompt that asked participants to pause and then explain how they knew that the headline of a Facebook news article was true or false. She found that doing so reduced the likelihood that people would share false news. However, the effect was lost on re-exposure to the same news items. Moreover, the stimulus was artificial: paper printouts of real and artificially faked versions of Facebook posts and prompts created by the experimenter. Bago, Rand, and Pennycook \cite{Bago2020} found that pausing increased the accuracy ratings of fake news that aligned with participants' political leanings. Later, Pennycook et al. \cite{pennycook2021shifting} explored having the user answer questions, aiming to prompt a shift in attention to the accuracy of news headlines/images. This increased the quality of the news shared. Also, people shared news that was not politically aligned with their own beliefs. While promising, this research was limited in the same way as that of Fazio \cite{fazio2020}. Chen, Ciuccarelli, and Colombo \cite{chen2022visualbubble} explored an indirect "pause and reflect" tool. Inspired by the "filter bubble" phenomenon, they created a browser extension for Facebook called \emph{VisualBubble}. The extension provided posts that were similar and different in opinion to that of individual users, i.e., in or outside their bubble. Some became skeptical but were not able to distinguish "false" news in general. Still, based on its potential, we incorporated the "pause and reflect" strategy in the \emph{Consider} form of our agent. We used real Tweets and a live Twitter setting with an agent rather than a human experimenter. We also compared this strategy directly with the "cover and warn" or \emph{Cover} strategy and the \emph{Camouflage} strategy, alongside a control group.

\subsection{Anti-Misinformation and Fact-Checking}
Sharing the results of fact-checking with users has also been explored. Vraga, Kim, and Cook \cite{Vraga2019} considered what style of corrections would be most effective: logic-based or humor-based. They found that success depended on the topic, with logic-based corrections being the most effective overall. Epstein, Pennycook, and Rand \cite{epstein2020will} considered a crowdsourcing mechanism of aggregating laypeople's evaluations of the trustworthiness of sources on social media. They found that people trusted mainstream sources more than hyper-partisan or fake news sources, indicating that crowdsourcing might be a viable fact-checking option. Diakopoulos, Trielli, and Lee \cite{diakopoulos2021towards} developed and investigated an AI-powered tool to draw journalists' attention to data that was subjected to quality and newsworthiness evaluations, but this was not assessed with lay users. 
In short, fact-checking seems to be an acceptable strategy for a variety of users.

Nevertheless, live fact-checking remains a technical challenge. While we designed our agent, Elemi, to use fact-checking as a basis for decision-making about misinformation, we were unable to develop a real-time fact-checking system---a project in its own right. Moreover, we wished to maintain a degree of experimental control, i.e., to compare equal sets of real and fake content. We also had concerns, emphasized by recent research \cite{ren2023beyond}, about whether people would be comfortable sharing their true thoughts and feelings about misinformation from their own feeds, i.e., content they or those in their social network produce or share. Due to these technical, methodological, and social challenges, we opted to curate and manually fact-check the Tweets used by our agent. 
We discuss this more in 3.1 and 3.2.

\section{Pilot Study of the \emph{Cover} Strategy: Explainable and Controllable Filtering}

We conducted a pilot study of the first version of our agent, Elemi, which used the \emph{Cover} strategy. Elemi hid Tweets containing misinformation, provided an explanation for the its decision based on fact-checking, and maintained user autonomy by allowing access to the content. We aimed to verify the positive results found in Sharevski et al. \cite{sharevski2021misinformation} with real rather than artificial content, increase efficacy with explanations based on fact-checking \cite{Vraga2019, epstein2020will,diakopoulos2021towards,ren2023beyond}, and determine the usability and UX of the agent within the actual context of use: live Twitter.
We asked: \emph{(RQ1) How user-friendly is the agent and what kind of experience does it provide?} and \emph{(RQ2) How effective is the "cover" strategy when deployed alongside a transparent, explainable AI system characterized by fact-checking and user autonomy in accessing hidden content?} We explored these questions in a user study format with direct evaluation and interview phases.

\subsection{System Design}

\begin{figure*}[h]
    \centering
    \includegraphics[width =\textwidth]{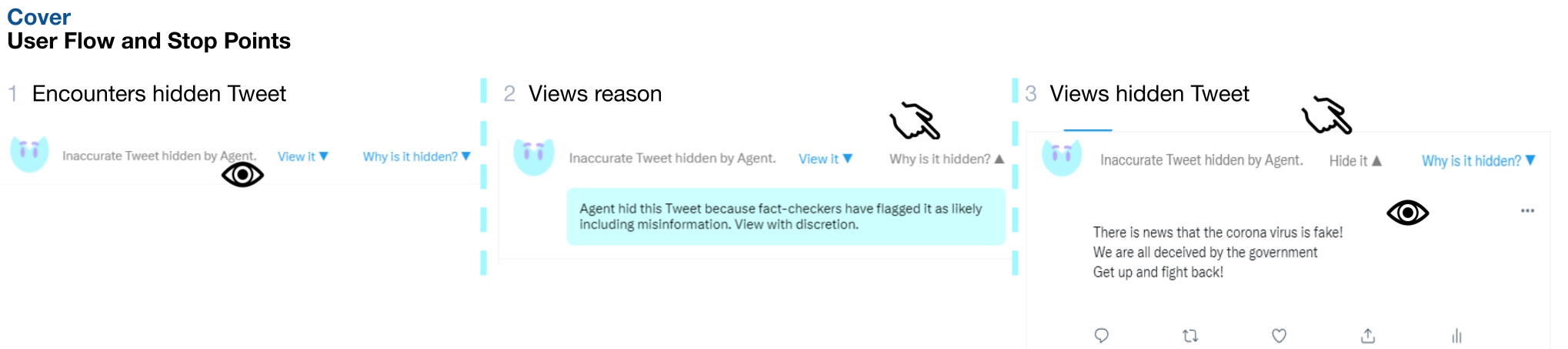}
    \caption{Overview of the \emph{Cover} version of the agent. Tweets with misinformation are hidden automatically by the agent (1). Users can access an explanation from the agent about why the tweet was hidden (2). They can also manually view the hidden tweet (3).
    }
    \label{fig:study1}
\end{figure*}

Our agent Elemi (Figure \ref{fig:study1}) was a browser-based tool designed in and for Twitter. It did not use the Twitter API, so it was not subject to the ongoing shifts in restrictions placed on third party apps\footnote{\url{https://www.engadget.com/twitter-new-developer-terms-ban-third-party-clients-211247096.html}}. The first version of the agent, \emph{Cover}, deployed the "cover" strategy of hiding tweets containing misinformation in a transparent and explainable way. The agent provided indicators that this was done and explained why with reference to fact-checking sources.

Elemi was designed to use fact-checking resources, but as a research prototype, it relied on the manual curation and fact-checking of tweets. 
When enabled, Elemi appeared as a cat-like, blue character icon on hidden tweets deemed to contain misinformation. The user could toggle the visibility of the tweet (user autonomy) and access an explanation about how Elemi based its decision on the results of fact-checking (transparency). 
Two researchers compiled the tweets used by the agent; refer to 3.2.3.
Being semi-autonomous, Elemi automatically labelled tweets within the live environment of Twitter, without direct human assistance. This was concealed from participants so as to simulate how it would actually work, i.e., without direct human involvement, allowing for a more accurate evaluation of the UX.

\subsection{Methods}

We conducted the pilot test on the \emph{Cover} version of Elemi between December 8\textsuperscript{th} to 23\textsuperscript{rd}, 2021.

\subsubsection{Participants}
Seventeen Japanese participants (13 men, 4 women, no one of another gender identity; aged 22-62) who had used Twitter at least once participated in the research. 
The research protocol was approved by the university ethics committee (approval \#2022170). All participants signed a consent form after receiving an explanation of the research. They were given compensation of equivalent to roughly \$11 USD.

\subsubsection{Location}
We conducted the study online and offline, due to restrictions arising from the COVID-19 pandemic. The offline version was conducted through a computer in the lab. The online version used Zoom and the Quick Assist program available in Windows 10. This allowed the participant to remotely control the lab computer, which allowed us to maintain the same virtual environment online and off, especially in terms of the browser environment.

\subsubsection{Materials}
The pre-screened tweets were fact-checked and categorized as either fake or true. For this, two researchers searched for real tweets about COVID-19 posted on Twitter between March 2020 and October 2021. 26 keywords were used as search terms, e.g., Vaccine, Corona, Corona graphene, Vaccine, Vaccine poison, Vaccine mask, mRNA spike, mask germs, mRNA poison, mRNA vaccine, Corona Electromagnetic Wave, Vaccine Danger, etc. Factual tweets usually referenced source data from the government or research articles, while fake tweets tended to lack evidence or media fact-checking. This led to more manual fact-checking efforts on the misinformation side. Fact-checking involved reviewing government materials and media articles. Tweets that both agreed to be true or fake were selected. From these, 15 factual and 15 fake tweets were randomly selected for the agent's use.

\subsubsection{Measures, Instruments, and Data collection}
Qualitative and quantitative data were collected. Participants' voices and faces during the think-aloud protocol were recorded. Audio was transcribed by a researcher who spoke fluent Japanese.
The post-test questionnaire consisted of fifteen items using Likert scales and three open questions.
\emph{Perceived usefulness} (PU) was used as a measure of initial trust \cite{benbasat2005trust} and assessed through four items from the instrument by Zhou et al. \cite{wang2016effects} on a 7-point Likert scale. \emph{Usability} was assessed through the 10-item System Usability Scale (SUS) \cite{brooke1996sus} on a 5-point Likert scale. We created one Likert-style item to assess frequency of Twitter use (no longer a user, never used, used once a month, once a week, daily). The open-ended questions asked about likes, dislikes, and what could be improved about the agent. For the interview, we asked about experience with the agent, use of Twitter, thoughts about Elemi's design,  understanding of the agent's capabilities, and willingness to use the agent again or not. All items were presented in Japanese; all English scales were translated into Japanese by one of the researchers who spoke fluent Japanese and English, and back-translated by two or more other researchers whose native language was Japanese. 

\subsubsection{Procedure}
Participants were given a description of the study and asked for consent. Next, they practiced the think-aloud protocol \cite{ericsson1984protocol}. For this, they read cat-related tweets for ~5 mins. They then completed the main task using the think-aloud protocol to narrate their experience. For this, they read tweets from a timeline comprising the curated 15 factual and 15 fake tweets, while the agent was activated. 
Participants were not directed to engage with the agent; they were free to use it or ignore it. Afterwards, they answered a post-test questionnaire and participated in an interview. We then ended the session. Participants were thanked and compensated. The study took about one hour.

\subsubsection{Data Analysis}
We generated descriptive statistics and scored the measures according to the instrument instructions. We used a two-tailed Student's t-test to compare the usability and PU scores by willingness or unwillingness to use the agent.
For the qualitative data, we used reflexive thematic analysis \cite{braun2006using}.
An inductive approach was used in the absence of a predefined thematic framework. For this, one researcher read the corpus of data drawn from the think-aloud protocol, interview transcripts, and questionnaires three times.
The researcher then developed 41 codes, later categorized under eight high-level codes: (1) experimental environment, (2) user interface, (3) feedback on the agent's functions, (4) inferences about the agent's logic, (5) UX, (6) feedback on the content, (7) opinion on the agent's information, and (8) agent improvements.
The researcher then grouped these into three themes, in light of the two RQs. These were reviewed several times with another researcher for clarity and refinement.

\subsection{Results}
\subsubsection{Quantitative Results}
 Most participants identified the agent's role as an anti-misinformation tool correctly; however, two did not. These were explainable: a non-frequent Twitter user and a non-student who believed that the timeline itself was created by the agent. We used the data from the fifteen participants who \emph{did} understand the agent's role. On average, they opened 43\% of the tweets marked as misinformation. 
The average normalized score for the SUS was 66.91, indicating marginal usability \cite{bangor2008empirical}. There were no significant differences in usability scores between the eight participants willing to use the agent ($70.25\pm19.88$) and the seven who were unwilling ($62.14\pm10.15$). The mean PU score was 4.53 (out of 7). A significant difference between the willing ($5.10\pm0.65$) and unwilling ($3.71\pm0.96$) groups, \emph{t}(15) = 3.576, \emph{p} < .01, indicated that those willing to use it found it more useful, as expected.

\subsubsection{Qualitative Findings}
We now turn to a review of our inductive findings from the thematic analysis.

\emph{Theme 1: Desire for transparency}.
Sixteen participants (all but one) wondered about the logic behind the agent. They investigated author profiles, the presence of quotes, the tone of the content, slang use, and other indicators to understand how the agent identified inaccurate tweets. Nine participants found that the filtering was not in line with their criteria and raised it as a problem. P1: \textit{"Well, this tweet hasn't been deleted, but I honestly don't believe it."} P8: \textit{"He doesn't seem like a person who would be likely to publish inaccurate information, so I'm surprised that his tweet is inaccurate, and he's using foreign data in English, or a summary of a paper, which seems to contain credible information."} Seven participants also expressed concern that the agent was biased and used only official or anti-COVID-19 information. P17: \textit{"I'm a little scared that the information that the agent is letting me see or blocks is decided arbitrarily."}

\emph{Theme 2: Trust and reliance}. Even when the agent was perceived as mistaken, some participants still trusted its decision. P11: \textit{"Even if the information is correct and the agent is wrong, I'd still accept the agent's decision."}. In fact, four participants hesitated to trust some tweets, stating that they would follow the agent's decision without knowing why it was made. P1: \textit{"I may believe this tweet because it's not hidden by this tool."} P6: \textit{"It's easy to believe because the agent didn't hide the tweet."}. Ten responded that they would use the agent in future.

\emph{Theme 3: Orientations to visual filtering}.
Impressions of the visual filtering changed over time with continued use. Nine participants found that the more they used the agent, the more it attracted their attention. P1: \textit{"I want to see what the conspiracy theory is by looking at the hidden tweets."} P4: \textit{"If such radical content is hidden, this might be a cue to click it, so I thought it might be a ploy by someone who wants to spread such radical ideas."}. Five participants also noticed and appreciated that the agent prepared them to consider the content before they opened it. P11: \textit{"It was good to embody the point of stopping and thinking about whether or not it was fake information."}

Still, about half did not necessarily want the agent to take action on misinformation. Eight participants enjoyed reading misinformation, finding it humorous. P13 criticized the filtering: \textit{"It's fun to hear various opinions, and I gain something from them. We must to decide ourselves whether to accept a specific opinion or not. But this agent takes away such a process. It seems for me there's no difference between a person deeply immersed in conspiracy theories and users of the agent."}. One participant suggested that it would be better to make the right information stand out rather than hide the wrong information. P3: \textit{"I'd like the agent to comment on something rather than just excluding negative tweets."}.

\subsection{Discussion}
We now discuss our findings from the pilot study based on the research questions.

\subsubsection{Agent Usability and UX (RQ1)}
The agent was generally usable, based on the SUS score (66.91) \cite{bangor2008empirical} and qualitative feedback. Frequent Twitter users immediately understood the agent's purpose and the "one click" toggle generated trust. Still, the presence of the agent iself drew attention to the tweets containing misinformation, leading many to focus on these fake tweets rather than stay vigilant. This points to the Pandora effect \cite{hsee2016pandora}: a desire to resolve uncertainty even while knowing that it will have a negative effect. Similar to Pennycook, Cannon, and Rand \cite{pennycook2018prior}, the "fact-checked" explanations provided by the agent served to increase perceived credibility, regardless of factuality.
However, the unintentional, negative side-effects need addressing, perhaps with metacognitive scaffolds or a means of allowing the user to guess what the hidden content might be even while not displaying the content directly \cite{stevens2007predator}.

\subsubsection{Agent Effectiveness (RQ2)}
Observations showed participants skipped nearly half of the content containing misinformation. In this sense, the agent was effective at preventing exposure to misinformation in live Twitter. But this experience was not pleasant for everyone. PU was linked to a desire to use the agent in the future. Some participants wished for a greater understanding of how the agent made decisions, while some were satisfied even when they thought that the agent made a mistake. Information that lacks transparency and a clear objective is unacceptable, even when true \cite{hartwig2019}. Others wished to view misinformation regardless, which may defeat the purpose of hiding it due to the threat of repeated exposure, where familiarity breeds a sense of accuracy, i.e., an illusory truth effect \cite{hassan2021effects}.

Overall, most people found the surface-level explanations about fact-checking to be insufficient when the agent covered up content. We also needed to tackle the challenge of repeated exposure. We thus returned to the literature, drawing on theories of metacognition as a means of assisting users who wish to engage with content containing misinformation directly. We developed two new versions of the agent, each representing a high-level strategy suggested as promising by previous work: \emph{Camouflage} \cite{stevens2007predator}, a new form of \emph{Cover} that blurs out (rather than hides) the content and specifies what parts were misinformed and why, as a metacognitive cue; and \emph{Consider}, which uses the metacognitive "pause and reflect" strategy investigated in numerous studies \cite{fazio2020,Bago2020}. We report on the study next.

\section{Experimental Study: The Metacognitive \emph{Camouflage} and \emph{Consider} Strategies}

We experimentally evaluated the new versions of Elemi, \emph{Camouflage} and \emph{Consider}, focusing on whether and how user behaviors and perceptions of misinformation changed with each strategy and over time. We used a between-subjects design with conditions for each version of the agent and a control condition (no Elemi, just regular Twitter). Two sessions were conducted two weeks apart to explore time-based effects, notably the \emph{illusory truth effect,} where prior exposure to fake news can increase perceptions of accuracy and willingness to share that news \cite{pennycook2018prior}. As before, the study took place in a lab setting but used real content in live Twitter. We asked: \emph{(RQ1) Does the \emph{Camouflage} strategy work better than the \emph{Consider} strategy at helping people identify misinformation?} Given our shift from hiding content to enabling a way of productively engaging with content using a metacognitive approach deployed over time,
we needed to check for an illusory truth effect. We asked: \emph{(RQ2) Does an agent employing a metacognitive approach mitigate the illusory truth effect and to what degree?} We begin by describing the new forms of Elemi, the agent.

\subsection{System Design}

We (re)designed the agent around the model of metacognition by Dunlosky and Metcalfe \cite{dunlosky2008metacognition}, focusing on the \emph{metacognitive strategies} component as a method users can employ when exposed to or choosing to engage with misinformation. This component, targeting the regulation of cognition, is made up of three activities: \emph{planning}, or deciding to be conscious of one's thinking and how to do so; \emph{monitoring}, or keeping track of thinking about one's thinking in the moment; and \emph{evaluating} one's success at the thinking task. In designing the agent to elicit certain activities, we adapted the framework of behavioural and cognitive interventions by Kozyreva et al. \cite{kozyreva2020citizens}. We explain below.

\begin{figure*}
\centering
\begin{subfigure}{\textwidth}
\includegraphics[width=\linewidth]{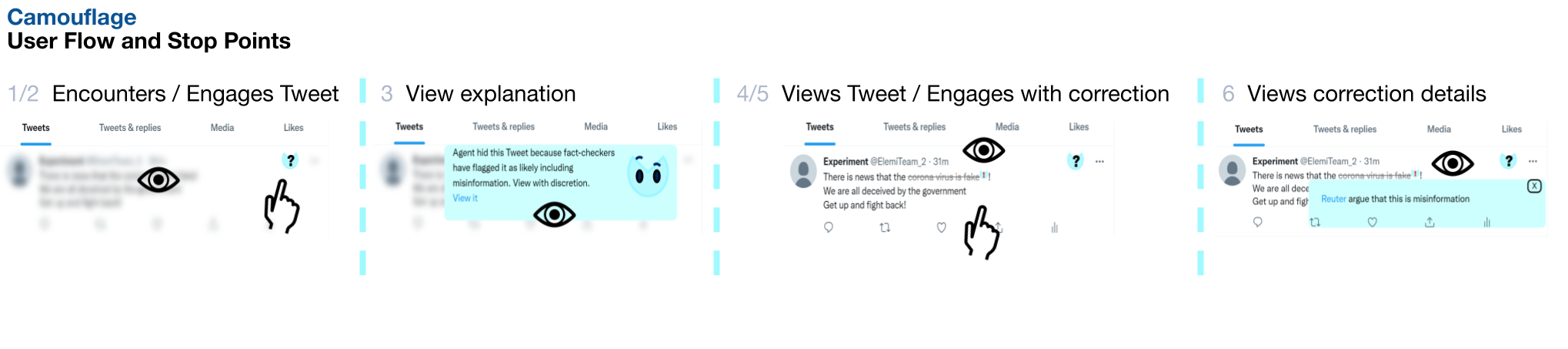}
\caption{Camouflage}\label{a--1}
\end{subfigure}\qquad

\begin{subfigure}{\textwidth}
\includegraphics[width=\linewidth]{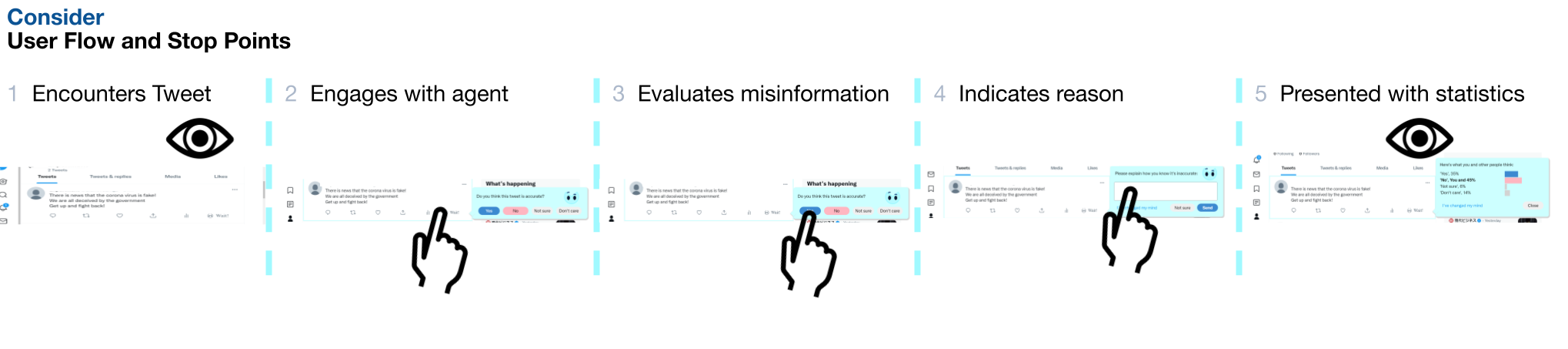}
\caption{Consider}\label{b--1}
\end{subfigure}\qquad
\caption{User flow for (a) \emph{Camouflage}: Tweets with misinformation are hidden (1). The user can engage the agent (2) for an explanation (3) and access the blurred out tweet (4). They can view the misinformation indicators (5) and fact-checking sources (6). User flow for (b) \emph{Consider}: The agent is next to the share button (1). The user clicks the agent and is asked for their opinion on the accuracy of the content (2). The user responds (3). The agent asks for the reason (4). The agent displays a graph of response data from other users (5).}
\label{fig:study2}
\end{figure*}

The \emph{Camouflage} version of the agent (Figure \ref{fig:study2}.a) restricted the content users were exposed to using a blur filter rather than hiding the content entirely, i.e., the \emph{Cover} strategy). Based on the pilot test, we expected this would act as a nudge \cite{kozyreva2020citizens}, but we could not predict in which direction. If the blurring effect reduced the visual saliency of the content, it could act as a nudge \emph{away} from the content \cite{stevens2007predator}. But if it drew attention as a novel visual element in the context of Twitter, it could nudge users \emph{towards} the content---a possibility that we could only evaluate through our novel experimental approach in live Twitter. The agent stimulated the planning stage of metacognition by explaining why the content was hidden, i.e., based on specific fact-checking sources, and allowing the user to unblur it. The agent also indicated where misinformation was detected in the text of the content with striked red text \cite{Vraga20202} and explanations on hover \cite{hartwig2019}. In essence, we aimed to improve the design of the \emph{Cover} strategy using a metacognition-inspired approach with a visual \emph{Camouflage} treatment and a more robust explanation, all the while retaining user control and transparency.

The new version Elemi, \emph{Consider} (Figure \ref{fig:study2}.b), offered a "technocognition" intervention \cite{kozyreva2020citizens} in the form of a "pause and reflect" strategy. Technocognition is about "directed friction," or encouraging deliberation of the content being consumed. This approach embodies the full metacognition cycle: planning, monitoring, and evaluating. An icon of the agent was placed next to the share button in each tweet, a critical zone in that UI; we aimed to divert attention from that button to the agent instead.
The icon drew attention by turning left and right at a $\pm5$ angle; its speech bubble also changed in size by 10\%. When clicking the agent, the user was asked about the accuracy of the content---prompting a planning activity---with four response options: \emph{accurate, inaccurate, not sure, don't care} \cite{pennycook2021shifting}. The agent then asked for the reason behind the response \cite{Bago2020}, prompting a monitoring activity. Then the agent provided a graph of other users' choices\footnote{The user choice data was predefined based on the fact-checking sources used by the agent \cite{Kim2020}.}, prompting an evaluation activity and completing the metacognitive cycle of engagement for that tweet.

\subsection{Methods}
We conducted a phased 2x3 controlled experimental study in which participants were pseudo-randomly assigned to one of three between-subjects conditions (\emph{Camouflage}, \emph{Consider}, or the control) and experienced the agent in two phases, roughly two weeks apart (within-subjects). We preregistered our methods and hypothesis with OSF\footnote{\url{https://osf.io/ucy9p}} on April 22\textsuperscript{nd}, 2022 and updated it on June 21\textsuperscript{st}, 2022.
The experiment ran from May 12\textsuperscript{th} to September 13\textsuperscript{th}, 2022.

\subsubsection{Participants}
We recruited people experienced with Twitter (Table \ref{Table:demographic}), with 57 in the first session and 49 completing the second. They were asked about their age and level of education after the second session to avoid stereotype threats \cite{kray2002reversing, mazerolle2012stereotype}. Participants were recruited through word-of-mouth, our social networks, and Prolific\footnote{\url{https://www.prolific.co/}}. They were compensated with at roughly \$11 USD per session. This study was approved by the university ethics committee (approval \#2022170).

\begin{table*}
  \caption{Age range and education level of participants}
  \label{tab:freq1}
  \begin{tabular}{ cccccl}
  \toprule
  Age Range & Overall & Control & Camouflage & Consider \\
  \midrule
  $18 \sim 24$ & 25 & 8 & 8 & 9\\
  $25 \sim 34$ & 17 & 6 & 5 & 6\\
  $35 \sim 44$ & 3 & 0 & 2 & 1\\
  $45 \sim 54$ & 2 & 1 & 1 & 0\\
  $55 \sim 64$ & 2 & 0 & 1 & 1\\
  \toprule
  Education Level &  &  &  &  \\
  \midrule
  High school degree or equivalent & 11 & 2 & 6 & 3\\
  Associate degree                 & 4 & 1 & 2 & 1 \\
  Bachelor's degree                & 14 & 5 & 4 & 5\\
  Graduate degree                  & 20 & 7 & 5 & 8\\
  \bottomrule
  Total & 49 & 15 & 17 & 17 \\
  \hline
  \end{tabular}
  \label{Table:demographic}
\end{table*}

\subsubsection{Location}
The study was conducted online and offline. Participants remotely controlled a lab computer in Zoom. This allowed us to control the study and ensure that all participants experienced the same environment.

\subsubsection{Materials}
The agent used the same fact-checking resources as in the pilot study, but the pre-screened tweets were changed to ensure up-to-date content. 
30 factual and 10 fake tweets posted between January and May 2022 were selected.
Unlike the pilot study, we allowed participation in either Japanese or English. For this, two researchers translated and back-translated the tweets. Also, all UI text and other content, such as ads and response tweets, were translated in realtime using the Google Translate API. English participants were notified that the original content was in Japanese to avoid confusion due to missed translations or oddities in the machine translation.

\subsubsection{Data Collection}
Tweet accuracy evaluations, misinformation exposure time, fact-checking activities, and usage logs were collected. Screen recordings with Zoom were used to capture observations of user activities. We verbally prompted participants to judge whether a tweet contained misinformation \cite{chen2022visualbubble}.
These assessments, as well as exposure time to misinformation content and other user activities, were transcribed from the recordings. The post-questionnaire used the same items as in the pilot study (refer to 3.2.4), in Japanese or English.

\subsubsection{Procedure}
There were two sessions, two weeks apart, with a different Twitter timeline each time that included 30 tweets: 20 factual and 10 fake. The 10 factual tweets changed between the first and the second session.
To start, participants were given a description of the study and asked for consent. They then carried out the main task, verbally speaking aloud their assessment of each tweet as they went. Unlike in the pilot study, they were asked to read the timeline to the end. 
Participants were not given any information of the agent and were free to engage with it or ignore it.
After the main task, they completed the post-questionnaire, and then were thanked and compensated.

\subsubsection{Data Analysis}
Descriptive statistics were generated. An individual accuracy score was obtained by dividing the number of accurate assessments by the number of tweets (0 to 1), one each per set of factual and fake tweets. We compared the time spent per fake tweets and during fact-checking activities by agent version and session to assess the effectiveness of each strategy and the presence of an illusory truth effect \cite{pennycook2018prior}. When groups contained less than five data points, Fisher's exact test was used instead of a Chi-square test. In addition, we used a Kruskal-Wallis test when comparing the accuracy score ratio, SUS, PU, and the time spent per fake tweet across the three conditions. For the qualitative data, 
we used an applied thematic analysis approach \cite{guest2011applied} with the goal of generalizing the findings by agent version and across participants. 
For this, two raters analyzed the data from the three open-ended questions.
The primary rater added on new themes and then each rater coded 20\% of the data separately. Inter-rater reliability was assessed with Cohen's kappa, with 0.8 as the criterion for agreement \cite{landis1977measurement}. Themes that did not meet this criterion were discarded or modified. Borderline cases were resolved by discussion. Similar themes were grouped together.

\subsection{Quantitative Results}

\begin{table*}
  \caption{Participants' fact-checking activities per session and agent version}
  \label{tab:freq3}
  \begin{tabular}{ccccc}
  \toprule
  & Category(Checklist) & Control & Camouflage & Consider\\
  \midrule
  \multirow{6}{*}{} & Author's profile & (11/19) & (12/19) & (13/19) \\
  \multirow{6}{*}{} & Twitter user reaction & (14/19) & (13/19) & (17/19) \\
  \multirow{6}{*}{} & Content attachment & (15/19) & (14/19) & (17/19) \\
  \multirow{6}{*}{} & Web searching & (5/19) & (6/19) & (7/19) \\
  \multirow{6}{*}{} & No fact-checking & (3/19) & (3/19) & (1/19) \\
  \bottomrule
  \multirow{6}{*}{} & Author's profile & (12/15) & (12/17) & (13/17) \\
  \multirow{6}{*}{} & Twitter user reaction & (12/15) & (17/17) & (14/17) \\
  \multirow{6}{*}{} & Content attachment & (12/15) & (15/17) & (17/17) \\
  \multirow{6}{*}{} & Web searching & (6/15) & (7/17) & (8/17) \\
  \multirow{6}{*}{} & No fact-checking & (2/15) & (0/17) & (0/17) \\
  \bottomrule
  \end{tabular}
  \label{Table:activity}
\end{table*}

\begin{table}
  \caption{Agent engagement frequency per session and agent version}
  \label{tab:freq4}
  \begin{tabular}{ ccccl }
  \toprule
  \multirow{2}{*}{Engagement} & \multicolumn{2}{c}{Session 1} & \multicolumn{2}{c}{Session 2} \\
  \multirow{2}{*}{} & Camouflage & Consider & Camouflage & Consider \\
  \midrule
  Skipped & 12 & 18 & 5 & 3\\
  Engaged & 7 & 1 & 12 & 14\\
  Active & 7 & 0 & 8 & 7\\
  \hline
  Total & 19 & 19 & 17 & 17\\
  \bottomrule
  \end{tabular}
  \label{Table:Freq}
\end{table}

\subsubsection{Influence of Agent Version on Fact-Checking Activities}
Participants engaged in fact-checking activities (Table \ref{Table:activity}). For all groups, the most frequent activity in Session 1 was checking "content attachments," followed by "Twitter users reactions" and "author profiles." Notably, those in the \emph{Consider} group engaged at the same frequency in "content attachment" and "Twitter users reaction" activities. One-third conducted "web searches" outside of Twitter.
In the second session, the trend was reversed. All in the \emph{Camouflage} group checked "Twitter users reactions," followed by "content attachments," "author profiles," and "web searching." All in the \emph{Consider} group checked "content attachments," followed by "Twitter users reactions," "author profiles," and "web searching." All in the agent conditions
performed at least one fact-checking activity in the second session. Regardless of condition, regular Twitter users actively sought out additional information before they judged whether the content was factual or fake.

\subsubsection{Influence of Agent Version on Perceived Accuracy of Tweets}
Accuracy scores for fake and factual tweets alongside exposure time to fake tweets per session and agent version are in Table \ref{Table:All}. 
In the \emph{Camouflage} group, seven (of 19) identified misinformation in the first session and twelve (of 17) did so in the second session.
These scores were compared by agent version and across time. No statistical significance was found for accuracy score or exposure time.

\subsubsection{Influence of Agent Version on Agent Engagement}
Agent engagement frequencies are in Table \ref{Table:Freq}. 
In the first session, seven (of 19) participants opened the \emph{Camouflage} agent's filter and read all false tweets. In the second session, twelve (of 17) opened up the filter and read the content at least once, and eight (of 17) uncovered nearly all content. In the \emph{Consider} group, one (of 19) interacted with the agent in the first session. Fourteen (of 17) in the second session used the agent at least once, and seven (of 17) interacted with it continuously. In the first session, the \emph{Camouflage} version of Elemi was interacted with more than the \emph{Consider} version (\emph{p} = .042), but not in the second session (\emph{p} = .688).

\subsubsection{Usability by Agent Version}
More than half of participants in Session 1 did not interact with Elemi As such, only the scores for Session 2 were compared. The SUS for \emph{Camouflage} was $5.24\pm1.14$ and for \emph{Consider} was $4.37\pm0.79$. There was a statistically significant difference found between \emph{Camouflage} and \emph{Consider}, \emph{t}(32) = 2.621, \emph{p} < .05, favouring \emph{Camouflage}. For PU, \emph{Camouflage} was $68.3\pm14.4$ and \emph{Consider} was $53.8\pm11.4$. There was also a statistically significant difference found, \emph{t}(32) = 3.250, \emph{p} < .01, indicating \emph{Camouflage} was preferred.

  
\begin{table*}
  \caption{Content evaluations for factual and fake content, exposure time, PU, and usability by session and agent version}
  \label{tab:freq5}
  \begin{tabular}{ clccccl }
  \toprule
   Session - Content & Measures & control & Camouflage & Consider\\
  \midrule
  \multirow{2}{*}{Session1 - Factual}& Number of participants engaged & 19 & 19 & 19 \\
  \multirow{2}{*}{}& Accuracy Score ratio $(0\sim1)$ & $0.75\pm0.16$ & $0.81\pm0.13$ & $0.76\pm0.17$\\
  \multirow{2}{*}{Session1 - Fake} & Number of participants engaged & 19 & 7 & 19 \\
  \multirow{2}{*}{}& Accuracy Score ratio $(0\sim1)$ & $0.85\pm0.15$ & $0.96\pm0.05$ & $0.87\pm0.14$\\
  \hline
  \multirow{2}{*}{Session2 - Factual}& Number of participants engaged & 15 & 17 & 17 \\
  \multirow{2}{*}{}& Accuracy Score ratio $(0\sim1)$ & $0.75\pm0.16$ & $0.74\pm0.19$ & $0.77\pm0.16$\\
  \multirow{2}{*}{Session2 - Fake} & Number of participants engaged & 15 & 12 & 17 \\
  \multirow{2}{*}{}& Accuracy Score ratio $(0\sim1)$ & $0.83\pm0.17$ & $0.81\pm0.34$ & $0.85\pm0.17$\\
  \hline
  \multirow{3}{*}{Session1} & Perceived Usefulness (PU) & - & $5.43\pm1.28$ & $4.90\pm1.35$ \\
  \multirow{3}{*}{}& System Usability Scale (SUS) & - & $72.5\pm14.2$ & $68.2\pm17.8$\\
  \multirow{3}{*}{}& Time spent per fake content (second) & $55\pm38$ & $49\pm26$ & $52\pm34$\\
  \multirow{3}{*}{Session2} & Perceived Usefulness (PU) & - & $5.24\pm1.14$ & $4.37\pm0.79$ \\
  \multirow{3}{*}{}& System Usability Scale (SUS) & - & $68.3\pm14.4$ & $53.8\pm11.4$\\
  \multirow{3}{*}{}& Time spent per fake content (second) & $60\pm47$ & $75\pm35$ & $69\pm36$\\
  \bottomrule
  \end{tabular}
  \label{Table:All}
\end{table*}


\subsection{Qualitative Findings}
 Here we present a summary of our thematic analysis findings, including a frequency analysis for each theme.

\subsubsection{Dis/misuse}
Use of Elemi varied by version, i.e., the metacognitive strategy it deployed. When it came to the \emph{Camouflage} version of Elemi, two participants were not aware of its existence. According to P40, \textit{"I wasn't even aware of the presence of an agent."} Among those who used Elemi, one (P18) expressed concern about dependence on the agent's judgments when in \emph{Camouflage} form: \textit{"I tend to judge by the content of the agent and less by the content of the tweets."}
In contrast, the \emph{Consider} version was intentionally not used by five participants. According to P50: \textit{"I dislike it because it just provides opinions, there's little objective about it."}

\subsubsection{Usability and UX}
Participants had various reactions to the user friendliness of the agent forms. Nine users of the \emph{Camouflage} version expressed that the agent was easy to use. As per P54: \textit{"It's lightweight and easy to learn."} Regarding its transparency and explainability, two user who never opened up its filter expressed positive opinions about its explainability. According to P41: \textit{"It's easy to use and integrated well with the social media platform."} Still, eight (of 12) who engaged with the agent expressed suspicion about its transparency and explainability. For instance, P13 wrote: \textit{"I doubt the authenticity of the agent's judgment ..."} Another advised that the \emph{Camouflage} version of Elemi may be a fit with other platforms, especially educational ones: \textit{"I think that detecting lies is a necessary skill when you grow up, so I feel it would be good to make it for children"} (P26). For two \emph{Consider} users, the agent was friendly. As per P8: \textit{"The design is catchy and friendly."} But when it came to ease-of-use, eleven (of 14) participants 
noted that the agent was not so easy to use. They all criticized the reasoning process. As P11 described: \textit{"I dislike having to go through each step ..."}

\subsubsection{Agent Performance}
Ten participants spoke to the \emph{Camouflage} version of Elemi. While six said that the accuracy of the fact-checking was high,
the other four indicated that some of the agent's judgments were incorrect. According to P16: \textit{"I feel that the sources for judging misinformation were somewhat scarce ..."} In short, more fact-checking was desired. At the same time, no comments were made about the \emph{Consider} version of the agent or other aspects of its performance.

\subsubsection{Dis/advantages}
Both versions of the agent were deemed good for decision-making support, i.e., as metacognitive tools. Still, the perceptions of Elemi's capability differed by version. Participants viewed the \emph{Camouflage} version of Elemi as an agent that confirmed the correctness or incorrectness of information. In contrast, \emph{Consider} was characterized as an agent that encouraged a rational decision: a different method of assistance. For \emph{Camouflage}, six participants commented that making incorrect information invisible would lead to platform comfort. According to P2: \textit{"The collection of tweets with credible sources was pleasant to see."}. Two said that the \emph{Camouflage} version of Elemi led them to be more aware of fake news. As per P33: \textit{"The warning made me more alert."}. Two others said that, on the contrary, their interest in the content of fake news rose. P31 explained: \textit{"Using an agent arouses interest in the fake news it's hiding."}
Five \emph{Consider} participants said that the agent inspired them to think logically. According to P7: \textit{"It helped me to judge logically rather than intuitively."} However, seven participants felt that there was a lack of objectivity in the agent's decisions. As per P52: \textit{"It's mainly or only opinion, there's little objective about it."} For P18, the \emph{Consider} version made it difficult to use the platform as casually as usual because it encouraged them to think about the information: \textit{..."I agree that fact-checking is important, but I felt that the value of Twitter as entertainment decreased ..."}

\subsection{Discussion}
We now discuss our findings based on the research questions.

\subsubsection{Does \emph{Camouflage} work better than \emph{Consider} at helping people identify misinformation? (RQ1)}

Leaving engagement with the agent up to the participant led to low engagement in the first session. Still, the accuracy score ratio of the \emph{Consider} version of the agent did not increase in the second session. As such, this strategy did not lead people to deliberate on the content as described by Bago, Rand, and Pennycook \cite{Bago2020}. As for \emph{Camouflage}, while it worked in the first session, its efficacy was disrupted in the second, apparently when tweets were repeated over time, i.e., repeated exposure, in terms of both accuracy score and user doubt. Transparency and clear explanations about the agent's decisions \cite{hartwig2019}, such as misinformation keywords and fact-checking references, did not prevent perceptions of the performance of the \emph{Camouflage} version of the agent from degrading on re-exposure to the same misinformation.

A phased, time-based approach to assessing the agent revealed the weakness of the \emph{Camouflage} version, indicating that it would not be superior to the \emph{Consider} strategy in the wild, at least in terms of helping people distinguish between factual and misinformed tweets to which they were exposed multiple times.
Indeed, neither \emph{Camouflage} nor \emph{Consider} alone were ideal as an anti-misinformation strategy over time. \emph{Camouflage}, which provided fact-checking data from the agent, was increasingly questioned by users from session to session. Given that agent behaviour can make users skeptical \cite{chen2022visualbubble}, repeated usage could increase the chance of users encountering inconsistencies with agent decision-making. These inconsistencies could exacerbate perceptions of distrust towards agent performance. \emph{Consider}, which required cognitive effort and made people think for themselves, was described as hard to use and inconvenient, having marginal acceptability and decreasing the entertainment value of social media. \emph{Consider} users also wanted objective information to support their decision-making rather than engaging in reasoning activities while reading the content. In the absence of objective feedback, they rated the agent's usefulness as poor, equivalent to the level of the "will not use" group in the pilot study. In Twitter, where content is instantaneous and unchallenged \cite{starbird2010chatter}, taking on cognitive load was opposed by participants. Therefore, a more practical approach could be to combine strategies, harmonizing the two agent versions as one and potentially including other strategies that we did not assess.






\subsubsection{Does an agent employing a metacognitive approach mitigate the illusory truth effect and to what degree? (RQ2)}

Changes in participants behaviour were captured over the course of two sessions. 
Unlike in previous research \cite{pennycook2018prior, hassan2021effects}, the illusory truth effect was not found.
Still, a notable difference was found in the \emph{Camouflage} group. Although it did not reach statistical significance as in Chen et al. \cite{chen2022visualbubble}, the \emph{Camouflage} group tended to be more skeptical than those in the \emph{Consider} and control groups over time. 
The metacognitive cue that \emph{Camouflage} used, blurring certain tweets as an indicator of misinformation and thereby implying that the rest was factual, was questioned by participants. The more inconsistent it was felt to be over time, the more that participants felt distrust towards the agent. Some even felt that the agent was misguiding them on purpose. The number of participants who disagreed with the agent doubled from the first to the second session (2 to 4). We might call this an \emph{amplified disbelief effect}. In short, those who had doubts about the agent's performance the first time were inclined to feel twice as doubtful the second time around. Thus, there is a chance that those who mistrust an agent from the start will further mistrust the agent in subsequent sessions.

\section{General Discussion}
This research aimed at empirically evaluating an "exoskeleton for the mind" in the form of Elemi, a metacognitive agent deploying several anti-misinformation strategies in the live context of Twitter. As a first step, we drew from existing work to create a \emph{Cover} version of the agent that filtered content. Pilot testing led us to improve this design and extend it with theories of metacognition applied to the digital realm of social media. \emph{Cover} became \emph{Camouflage}, which we empirically evaluated in a phased experimental study alongside a new version of the agent, \emph{Consider}. This version deployed a "pause and reflect" strategy that captured the full cycle of theorized metacognitive activities. We compared each version to a control. Overall, we found no statistically significant differences in terms of participant activities, misinformation exposure time, and accuracy score rating. However, people rated the \emph{Camouflage} version of Elemi much higher than its predecessor and its counterpart, \emph{Consider}. We also did not find evidence of an illusory truth effect, although we did find evidence of another effect for those who claimed reason to distrust the agent from the start: what we call \emph{amplified disbelief effect}. We discuss the implications below.

\subsection{Still In Search of Strategies}
Classic countermeasures for misinformation, namely warning and filtering, are deployed based on the severity of the content in question. Yet, in modern social media contexts, countless people become sources of information on the fly, making it difficult to identify and respond to misinformation in a timely and effective manner \cite{grinberg2019fake}. Automated methods, such as Meta/Facebook's approach\footnote{\url{https://www.facebook.com/formedia/mjp/programs/third-party-fact-checking}}, are often deployed, but such methods are known to be slow \cite{pennycook2018prior,fazio2020} and require trust in the company. Likewise, the \emph{Camouflage} version of our agent revealed cases where users questioned the agent's filtering and had doubts about whether the filtered information was misinformation, thus inspiring direct engagement with the content the second time around. This \emph{amplified disbelief effect} decreases the utility of such agents, even biasing user decisions towards misinformation. A counter strategy is needed to reap the rewards of a \emph{Camouflage} strategy.

However, this does not necessarily mean that user-driven strategies work better than the agent-driven ones. Several anti-misinformation strategies and psychological studies on the "pause and reflect" approach have shown high effectiveness for user vigilance towards fake news \cite{Bago2020, fazio2020}. However, it seems that inculcating "metacognitive behaviour" into social media platforms requires a number of hurdles to be overcome. The scoring ratio patterns between the control and \emph{Consider} conditions in our work indicate that forcing deliberation within the social media context went nowhere \cite{Bago2020}. The \emph{Consider} version of Elemi was rated as unfit for a casual social media context, and forced cognitive engagement was tiresome. Future anti-misinformation and/or metacognitive agent designs should take into account how to motivate users. As we confirmed, when given leave to act freely, Twitter users were uncooperative, especially when it came to reasoning activities and even under the observation a researcher. This implies that they could do worse when alone.

These findings suggest that anti-misinformation and/or metacognitive strategies should be integrated into agents in a way that complements user needs and expectations with the given social media platform. Still, hiding fake content, using either a \emph{Cover} or \emph{Camouflage} strategy, was particularly useful for preventing exposure to false information, at least at first. Positive feedback alongside high PU and usability scores were maintained as long as the filtering allowed for user autonomy and accessibility. Still, we need to find a way to motivate users to not open the filter needlessly. One example could be combining a user-driven evaluation approach, such as letting users rate on the first encounter, and then reminding them of this rating on subsequent exposure to the content, ideally with supporting articles and sources for fact-checking purposes. This would also address user feedback on the \emph{Consider} form of the agent, in terms of a desire for more objective feedback. The agent should certainly prepare sufficient evidence as to why a given piece of content is fake (or not) before users are asked to judge that content, so as not to generate suspicion.

Another point to bear in mind is when to have the agent help the user confront misinformation. Chen et al. \cite{chen2022visualbubble} showed that users became more skeptical of information while iterating through judgments of whether the information was true or false. This was also found for the case of the \emph{Camouflage} version of Elemi. In this respect, the role of \emph{Camouflage} is limited. Judgments involve cognitive load, so user-driven methods should consider how to exercise cognition in a way that leads to metacognition activities that are effective and not overwhelming.

\subsection{Implications for Design}

We now summarize our findings in a set of general design implications for similar in situ anti-misinformation tools that aim to include elements of metacognition:

\begin{itemize}
  \item Hidden misinformation should still be accessible to users for reasons of trust, autonomy, and transparency.
  \item Filtering should be accompanied by clear scaffolding to prevent repeated access to the misinformation, or at least contextualize the user's previous reactions to that content, such as by providing the user's original ratings.
  \item A repeated mismatch between user and agent judgments and decisions on the misinformed nature of content or fact-checking activities could induce an \emph{amplified disbelief effect} that biases the user towards, rather than away from, misinformation. Counter strategies, personalization, and reliance on objective sources should be explored.
  \item Forced cognitive engagement may not be welcome by most users, especially regular Twitter users.
  \item How to engage users, especially in consideration of cognition load within the casual context of many social media platforms, should be considered and tested with respect to helping users remain vigilant to misinformation. Users should be guided towards metacognitive activities, if possible.
  
\end{itemize}

\subsection{Limitations}
Not all participants completed the second session, limiting the time-based results. Additionally, most participants were students in engineering, so future research should recruit a more diverse sample. It is also possible that the topic of COVID-19, a known polarizing topic, influenced impressions of the agent itself. Future work can include different topics, hashtags, and tweets. We also made important decisions about the design of the agent and the research design, notably opting for a balance of realism and experimental control in light of practical limitations with the prototype. Our findings indicate that this was beneficial in allowing participants to feel comfortable and freely provide their opinions, as well as revealing unusual results, especially over time with the nullified illusory truth effect. Nevertheless, future work should fully automate the agent with a live fact-checking service and re-test with user's real accounts and timelines.

\section{Conclusion}
Anti-misinformation tools are on the rise. Many strategies and ideas have been tested experimentally, primarily outside of the wilderness of social media platforms like Twitter, with middling or unclear results. We aimed to increase the ecological validity of this body of work and directly compare the efficacies of promising strategies. Our work on Elemi shows that such tools may need special care to work in live social media environments with a range of potential end-users. Future work can explore the design implications proposed with diverse personalities on a greater range of platforms in an autonomous form distributed and tested in live environments with non-curated content.

\begin{acks}
This work was funded by a DLab Challenge: Laboratory for Design of Social Innovation in Global Networks (DLab) Research Grant (Tokyo Institute of Technology). We thank the Aspirational Computing Lab for early feedback.
\end{acks}

\bibliographystyle{ACM-Reference-Format}
\bibliography{References}

\end{document}